\documentclass[proof]{pasj00}
\Received{$\langle$reception date$\rangle$}
\Accepted{$\langle$acception date$\rangle$}
\Published{$\langle$publication date$\rangle$}
\SetRunningHead{Inoue et al.}{Ice feature in scattered light}

\usepackage{times}

\begin{document}

\title{Observational Possibility of the ``Snow Line'' on the Surface
of Circumstellar Disks with the Scattered Light}
\author{Akio K. \textsc{Inoue}}
\affil{College of General Education, Osaka Sangyo University, 
3-1-1, Nakagaito, Daito 574-8530}
\email{akinoue@las.osaka-sandai.ac.jp}
\author{Mitsuhiko \textsc{Honda}}
\affil{Department of Information Science, Kanagawa University,
2946 Tsuchiya, Hiratsuka, Kanagawa 257-0031}
\email{hondamt@kanagawa-u.ac.jp}
\author{Taishi \textsc{Nakamoto} \and Akinori \textsc{Oka}}
\affil{Department of Earth and Planetary Sciences,
Tokyo Institute of Technology, Ookayama, Meguro, Tokyo 152-8551}
\email{nakamoto@geo.titech.ac.jp, akinori@geo.titech.ac.jp}


\maketitle

\begin{abstract}

We discuss how we obtain the spatial distribution of ice on the 
surface of the circumstellar disk around young stars.
Ice in the disks plays a very important role in various issues, 
for instance, on the disk structure, on the planet formation, 
on the isotopic anomaly in meteorites, and on the origin of 
the sea on the Earth.
Therefore, the spatially resolved observation of the 
condensation/sublimation front of ice, 
so-called ``snow line'' is strongly required.
Here, we propose a new method for obtaining the spatially resolved 
``snow line'' on the circumstellar disks by observing 3 \micron\ 
H$_2$O ice feature in the scattered light.
Based on radiative transfer considerations, 
we show that the feature is clearly imprinted in the spectrum 
of the scattered light from both optically thick and thin 
circumstellar disks. 
We also show that the scattered light and the H$_2$O ice feature 
from protoplanetary disks 
are detectable and spatially resolvable with the current instruments 
through a $H_2O$ narrowband filter around 3 \micron. 
Finally, we present a diagnostics of disk dust properties 
on the $K-H_2O$ and $K-L'$ two color diagram.

\end{abstract}

\section{Introduction}

Ice of water plays many important roles in the protoplanetary disk.
First, ice enhances the opacity and affects the disk structure 
(e.g., \cite{lin80,pol94,oka07}). 
Second, ice enhances the amount of solid material in the disk, 
which promotes the formation of protoplanets and
the core of gaseous planets
(e.g., \cite{hay85,kok02}).
Third, ice evaporation in the inner region of the disk may cause 
the isotopic anomaly of oxygen found in meteorites \citep{yur02}.
Fourth, icy planetesimals or comets may bring water on the Earth 
\citep{mor00,ray04} as well as the innermost part of the disk 
\citep{eis07}.
To discuss these issues in detail, we need the spatial distribution 
of ice in the disk. 
Therefore, we should resolve the ice condensation/sublimation front,  
so-called ``snow line'', in the disk.

Ice is found ubiquitously in the molecular cloud and the circumstellar
envelope through the prominent 3 \micron\ absorption feature 
(e.g., \cite{gib04}). In two edge-on protoplanetary disks, the 3
\micron\ ice absorption feature is clearly detected by spectroscopy with
the Subaru telescope \citep{ter07}. 
However, we cannot resolve the ``snow line'' along the line of 
sight with this method. Observations with the Infrared Space Observatory
found 44 and 62 \micron\ ice emission feature in some protoplanetary
disks \citep{mal99}. However, very low spatial resolution of the
far-infrared instrument prevents us from resolving the ``snow line'',
either. 

On the other hand, 
the scattered light from edge-on protoplanetary disks have already been
detected well in the optical to the mid-infrared by very high-resolution
imaging with the Hubble Space Telescope and ground-based 8-m telescopes 
(e.g., \cite{bur96,kor98,mcc03}). 
The scattered light from debris disks have also been detected in the
optical to the near-infrared (NIR) (e.g., \cite{smi84,gol06,kal07}).
Interestingly, \citet{gol06} suggest that the existence of icy grains is
a possible cause of the observed reddening of the optical scattered
light beyond $\sim120$ AU in the debris disk of $\beta$ Pictoris. 
However, \citet{gri07} argue that the photosputtering by ultraviolet
photons results in the ice erosion in optically thin disks. To resolve
this issue, we should seek a direct evidence of the ice if it exists.

The most important merit of the scattered light observations is the 
ability to reveal the detailed surface structure of the disks.
Indeed, coronagraphic imaging with the Subaru telescope has detected the
scattered light from face-on disks in the NIR 
and revealed the surface structure \citep{fuk04,fuk06}. If an evidence of
ice is imprinted in the scattered light, we can resolve the distribution
of ice, i.e. ``snow line'', with the scattered light. In section 2, 
we first show that the H$_2$O ice absorption feature at 3 \micron\ is
clearly imprinted in the scattered light spectrum, and then, we present
a few diagnostics diagrams to check the existence of ice as well as a
typical grain size. In section 3, we discuss the observational
feasibility of the icy grains with the scattered light. The final
section is a summary of this paper.

\section{Ice feature in the scattered light}

\subsection{Observable scattered light}

We consider the radiation transfer of only the light scattered by the
circumstellar disk. 
That is, we consider a wavelength where the thermal radiation from the
disk is negligible.
Let us consider a ray toward an observer as shown in Fig.~1.
We set an optical depth coordinate $t$ on the ray. 
For simplicity, we assume that 
the disk has a clear ``surface''. In other words, 
there is no matter in the outside of the disk. 
We set the total optical depth of the disk along the ray at the
frequency $\nu$ as $\tau^{\rm d}_\nu$. 
Thus, the optical depth coordinate has $t=0$ and $t=\tau^{\rm d}_\nu$ at
the two ``surfaces'' of the far and the near sides from the observer,
respectively (see Fig.~1). Note that the optical depth 
$\tau^{\rm d}_\nu = (\kappa^{\rm a}_\nu + \kappa^{\rm s}_\nu) \Sigma$
with the absorption and scattering cross sections per unit gas mass 
$\kappa^{\rm a}_\nu$ and $\kappa^{\rm s}_\nu$, respectively, and 
the gas mass column density along the ray $\Sigma$. 
We will assume a typical mass abundance of dust relative to gas based
on the elemental abundance of the Sun later. 
The scattering albedo 
$\omega_\nu=\kappa^{\rm s}_\nu/(\kappa^{\rm a}_\nu+\kappa^{\rm s}_\nu)$.
 
The ``surface'' of the real circumstellar disks can be defined by
various ways; the photosphere of the dust emission, the scale height of
the material distribution, and the layer at which the optical depth
against the stellar radiation becomes unity. 
For the context of this paper, the last two definitions could be
realistic for optically thin and thick disks, respectively. 
However, we here assume a simpler ``surface'' in which all the
materials are confined. This assumption might be too simple, but we can
highlight the relevant physics clearly. In addition, the results from
this assumption excellently agree with those from more complicated
numerical simulations as shown below. 

The observable intensity without the incident radiation along the ray is 
\begin{equation}
 I^{\rm obs}_\nu = \int_0^{\tau^{\rm d}_\nu} 
  S_\nu(t) e^{-(\tau^{\rm d}_\nu-t)} dt, 
\end{equation}
where $S_\nu(t)$ is the scattering source function along the ray 
at the optical depth coordinate $t$ shown in Fig.~1.
We consider a point P($t[R,z]$) in the disk (see Fig.~1). 
Assuming the isotropic scattering for simplicity, we have 
\begin{equation}
 S_\nu(t[R,z]) = \omega_\nu J^*_\nu(R,z) + \omega_\nu J_\nu(t)\,,
\end{equation}
where $J^*_\nu$ is the mean intensity of the stellar radiation; 
\begin{equation}
 J^*_\nu(R,z)=\frac{1}{4\pi}B_\nu(T_*)\Omega_*(R,z)e^{-\tau^*_\nu(R,z)}\,,
\end{equation}
where $B_\nu(T_*)$ is the Planck function with the stellar effective
temperature $T_*$, $\Omega_*$ is the solid angle of the stellar
photosphere from the point P, and $\tau^*_\nu$ is the optical depth
between the stellar photosphere and the point P.  
The solid angle is 
$\Omega_*=\pi f_{\rm vis}{R_*}^2/(R^2+z^2)$ with the stellar
radius $R_*$ and the visible fraction of the stellar photosphere from
the point P $f_{\rm vis}$. 
If we consider a ray almost face-on and assume a disk geometrically
thin, the solid angle $\Omega_*$ can be regarded as constant; 
$\Omega_*=\pi f_{\rm vis} (R_*/R)^2$. 
The second term in eq.~(2) is due to the multiple scattering in the
disk; $J$ is the mean intensity of the scattered radiation at
the point P.

In the optically thin disk, we can consider $\tau^*_\nu\approx0$ 
and can neglect the multiple scattering; $J\approx0$. 
Thus, we have 
\begin{equation}
 S_\nu(t) \approx S^*_\nu 
  \equiv \omega_\nu B_\nu(T_*) \frac{\Omega_*}{4\pi}\,,
\end{equation}
which can be regarded as constant along the ray if $\Omega_*$ is
approximated to be constant. 
In this case, we can integrate eq.~(1) easily, and we have 
\begin{equation}
 I^{\rm obs}_\nu = S^*_\nu (1-e^{-\tau^{\rm d}_\nu})
  \approx \tau^{\rm d}_\nu S^*_\nu\,,
\end{equation}
where we have applied the condition $\tau^{\rm d}_\nu \ll 1$ of  
an optically thin disk. 

For the optically thick disk, we can consider 
$\tau^*_\nu \approx (\tau^{\rm d}_\nu-t)/\beta$, 
where $\beta$ is so-called grazing angle; the angle between 
the incident radiation and the disk surface (see Fig.~1). 
Generally, $\beta$ is as small as about 0.05 radian. 
Although $\beta$ increases along the radial distance $R$ from the
central star in a flaring disk, we expect $\beta \lesssim 0.1$ radian
even at $R \gtrsim 100$ AU \citep{dal06}. 
For a single scattering case, we have 
\begin{equation}
 S_\nu(t) \approx S^*_\nu e^{-(\tau^{\rm d}_\nu-t)/\beta}\,,
\end{equation}
where $S^*_\nu$ is defined in eq.~(4). In this case, we can integrate
eq.~(1) with the assumption of constant $\beta$ and 
$1+1/\beta \approx 1/\beta$, then, 
\begin{equation}
 I^{\rm obs}_\nu \approx \beta S^*_\nu (1-e^{-\tau^{\rm d}_\nu/\beta})
  \approx \beta S^*_\nu\,,
\end{equation}
where we have applied the condition $\tau^{\rm d}_\nu \gg 1$ of  
an optically thick disk. However, the multiple scattering could be
important for the optically thick disk. According to the ``law of
diffuse reflection'' by semi-infinite plane-parallel medium 
\citep{cha60}, this effect is accounted for by the $H$-function; 
\begin{equation}
 I^{\rm obs}_\nu(\mu) \approx \beta S^*_\nu H(\mu,\omega_\nu)\,,
\end{equation}
where $\mu$ is cosine of the reflection angle from the normal of the
slab. We have adopted an approximation with $\beta\approx0$ and
$\mu\approx1$ (i.e. almost face-on). 
\footnote{The exact form of the ``law of diffuse reflection'' is 
(\cite{cha60}, sec.~III, eq.~118)
$$
 I^{\rm obs}_\nu(\mu) = \omega_\nu \frac{F_\nu}{4}
  \frac{\mu_0}{\mu+\mu_0}H(\mu,\omega_\nu)H(\mu_0,\omega_\nu)\,,
$$
where $F_\nu$ is the flux density incident on the plane-parallel slab
with the angle from the normal whose cosine is $\mu_0$. 
Note that $\mu_0=\sin\beta\approx\beta\approx0$, 
$\omega_\nu F_\nu/4=S^*_\nu$ as eq.~(4), and $H(0,\omega_\nu)=1$.}

Finally, we have the observable intensity toward the face-on direction
as follows: 
\begin{equation}
 I^{\rm obs}_\nu \approx \cases{ 
  \displaystyle \beta \omega_\nu H(1,\omega_\nu) 
  B_\nu(T_*) \frac{\Omega_*}{4\pi} 
  & (optically thick disk) \cr
  \displaystyle  \tau^{\rm d}_\nu \omega_\nu 
  B_\nu(T_*) \frac{\Omega_*}{4\pi} 
  & (optically thin disk) \cr} \,.
\end{equation} 
Note that 
$\tau^{\rm d}_\nu\omega_\nu=\kappa^{\rm s}_\nu\Sigma=\tau^{\rm s}_\nu$,  
where $\tau^{\rm s}_\nu$ is the scattering optical depth of the disk. 
Therefore, we can observe the dust feature via the albedo $\omega_\nu$ or
the scattering cross section $\kappa^{\rm s}_\nu$ (or $\tau^{\rm s}_\nu$) 
in the scattered light from circumstellar disks.

In Fig.~2, we show the scattering albedos and cross sections 
of silicate and silicate+ice dust as a function of the wavelength 
in the NIR for some cases of the grain size.
The optical properties are taken from \citet{miy93}; 
astronomical silicate \citep{dra85} and crystalline H$_2$O ice 
\citep{irv68,ber69,sch73}.
We calculated the absorption and scattering coefficients of the two
species independently by the Mie theory \citep{boh83} with the mass
abundances relative to gas of $\zeta_{\rm sil}=0.0043$ and 
$\zeta_{\rm ice}=0.0094$ and with the material density of 
$\delta_{\rm sil}=3.3$ g cm$^{-3}$ and $\delta_{\rm ice}=0.92$ g cm$^{-3}$ 
\citep{miy93}, then, we added them. Even if we calculated the
coefficients as ice-coated silicate or silicate-ice mixture, 
the presence of the H$_2$O ice features would not be affected
qualitatively \citep{miy93,pre93}.

As shown in Fig.~2, we expect in general a steeper spectral slope
(i.e. bluer color) as a smaller dust grain size because $\omega_\nu$ and
$\kappa^{\rm s}_\nu$ of smaller grains have a strong wavelength
dependence. Since $\omega_\nu$ and $\kappa^{\rm s}_\nu$ are almost
independent of the wavelength, i.e. gray scattering, for $\gtrsim1$
\micron\ grains, the spectral slope is expected to be very similar to
that of the Rayleigh-Jeans law in the NIR. 
In the case with H$_2$O ice, a strong absorption feature of H$_2$O ice
appears in the albedo at 3 \micron\ and other features, for example, of
2 \micron\ and 4.5 \micron, are also found, but weaker (panel [b]). The
3 \micron\ O-H stretching mode also shows the typical resonance feature
in the scattering cross section (panel [d]). The strength of the feature
seems to decrease as the dust grain size increases although it does not
disappear even in the 10 \micron\ case.

\subsection{Spectrum and color with ice}

Fig.~3 shows the expected spectra of the brightness of an annulus with
silicate and H$_2$O ice;  
the panel (a) is the optically thick case and the panel (b) is the thin
case. We have assumed the dust size of 1.0 \micron, the radius of the
annulus of 100 AU, and a central A-type star ($T_*=10,000$ K and
$R_*=2.5$ \RO). The gas surface density of the annulus is assumed to be 
$1.7$ g cm$^{-2}$ (e.g., \cite{hay85}) and $1.7\times10^{-5}$ g
cm$^{-2}$ for the thick and the thin cases, respectively. The grazing
angle $\beta=0.05$ and the visible fraction of the stellar photosphere
$f_{\rm vis}=0.5$ are also assumed.
The geometry between the observer and the annulus is face-on. 

The solid curves in Fig.~3 are calculated by eq.~(9). For the
optically thick case, we need the $H$-function. 
Here we adopt an approximate expression of it by \citet{bri88} rather
than the original one whose evaluation needs an iterative calculation; 
\begin{equation}
 H(\mu,\omega_\nu) \approx \frac{1+2\mu}{1+2\mu \sqrt{1-\omega_\nu}}\,.
\end{equation}
The dotted curves are results from numerical simulations; 1+1D radiative
transfer calculation with the variable Eddington factor method
developed by \citet{dul02} but extended by us to treat the isotropic
scattering (Inoue, Nakamoto, Oka in preparation). We show the spectra of
the single annulus same as the solid curves. In the calculation, we
kept $\beta=0.05$ but iteratively solved the vertical hydrostatic
equilibrium in the annulus for the central stellar mass $M_*=2.5$ \MO\
to be consistent with the temperature structure from the radiative
equilibrium. As shown in the both panels of Fig.~3, we find excellent
agreements between the simple analytic models of eq.~(9) and the
numerical simulations. 

In the NIR around the 3 \micron\ H$_2$O ice absorption
feature, the stellar spectrum can be expressed by the Rayleigh-Jeans law. 
Thus, as an approximation, the scattered spectra may be given by 
\begin{equation}
 I^{\rm obs}_\nu \propto \cases{
  \omega_\nu \nu^2 
  & (optically thick disk) \cr
  \kappa^{\rm s}_\nu \nu^2 
  & (optically thin disk) \cr } \,, 
\end{equation}
which are shown as the dashed curves in Fig.~3 (see also eq.~13). 
Indeed, we find that this approximation is good enough to discuss just
if the 3 \micron\ H$_2$O ice feature is observable or not. Hereafter, we
use this simple expression as the scattered spectra. 

Fig.~4 shows two photometric colors expected as a function of the dust
grain size: (a) and (b) for $K-H_2O$, and (c) and (d) for $H_2O-L'$. 
We have assumed a usual set of the filter transmission functions of
$K$ and $L'$.\footnote{We used the machine readable form downloaded from
the Subaru/CIAO web page; 
http://subarutelescope.org/Observing/Instruments/CIAO/camera/sensitivity.html}
This choice of the bands is not definitive. We could choose any two
bands shortward and longward of $H_2O$ band.

The $H_2O$ band transmission function is assumed to be a
rectangular between 3.02 and 3.16 \micron. The magnitude zero points 
are taken from \citet{coh92} for $K$ and $L'$. For $H_2O$, 
we assume $F_\nu{\rm (Vega)}=357$ Jy at $\lambda=3.09$ \micron\ 
based on Fig.~2 in \citet{coh92}. Four disk models, optically
thick disk with silicate (open circles), optically thick disk with
silicate and ice (filled diamonds), optically thin disk with silicate
(open triangles), and optically thin disk with silicate and ice (filled
squares), are shown in the panels. The dotted lines indicate the
Rayleigh-Jeans law.

The scattered light colors without H$_2$O ice (open circles and triangles) 
are bluer than the Rayleigh-Jeans law for a small ($\lesssim1$ \micron)
dust size\footnote{In this paper, we do not consider any specific size
distribution function of the dust particle, but consider a ``typical''
size of the particle. This ``typical'' size can be defined by an average
over the size distribution function with a certain weight, for example,
cross section in a certain band. We can consider a ``typical'' size of
the interstellar dust to be $\sim0.1$ \micron.}
and they approach the Rayleigh-Jeans color as the dust size
increases. The observational fact that most of the scattered light from
the disks are not bluer than the stellar color indicates a large
($\gtrsim1$ \micron) dust size (but AU Mic; \cite{kri05}).

If the dust includes H$_2$O ice (filled diamonds and squares), 
significant differences appear; for optically thick disks with ice
(diamonds), $K-H_2O$ is much bluer than that without ice and $H_2O-L'$
is much redder than that without ice, because of the prominent
absorption in $H_2O$ band. 
Note that $K-H_2O$ becomes blue even if a large ($\gtrsim1$
\micron) dust size which makes gray scattering in $K-L'$. 
Thus, we can conclude that there is H$_2$O icy dust with a large size if
we observe neutral $K-L'$ and blue $K-H_2O$ colors. On the other hand, 
we can conclude that there is H$_2$O ice only if we observe red
$H_2O-L'$ because the red color is not produced without H$_2$O ice.

For optically thin disks with ice (squares), 
$K-H_2O$ starts from the Rayleigh-Jeans color (i.e. much redder color than 
that without ice) and becomes gradually bluer as grain size becomes larger.
$H_2O-L'$ oppositely starts from a very blue color and becomes gradually 
redder as grain size becomes larger. Such behaviors are caused by 
the resonance feature in $H_2O$ band seen in the scattering cross
section (Fig.~2 [d]).

Interestingly, these differences caused by the feature in $H_2O$ band 
enable us to classify disk and dust properties on a two color diagram,
for example $K-H_2O$ and $K-L'$, as shown in Fig.~5.
In this figure, we find four sequences: the optically thick disk with
silicate (open circles), the optically thick disk with silicate and ice
(filled diamonds), the optically thin disk with silicate (open
triangles), and the optically thin disk with silicate and ice (filled
squares). Each sequence consists of five points corresponding to the
grain sizes of 0.1, 0.3, 1.0, 3.0, and 10 \micron. The large asterisk
indicates the location expected from the Rayleigh-Jeans law. In Fig.~5,
we find five classifications: 
\begin{description}
 \item[A:] optically thick and thin disks 
	    with small ($\lesssim1$ \micron) silicate dust 
 \item[B:] optically thick and thin disks 
	    with large ($\gtrsim1$ \micron) silicate dust
 \item[C:] optically thick disks 
	    with small ($\lesssim1$ \micron) silicate and ice dust 
 \item[D:] optically thick and thin disks 
	    with large ($\gtrsim1$ \micron) silicate and ice dust
 \item[E:] optically thin disks 
	    with small ($\lesssim1$ \micron) silicate and ice dust 
\end{description}
If we observe a disk through 3 bands, $K$, $H_2O$, and $L'$, then, 
we plot the data on the two color diagram, we can judge if the disk 
is optically thick or thin, the grain size is smaller or larger than 
about 1 \micron, and the dust includes ice or not.
Importantly, if we plot spatially resolved data on the diagram, 
we can derive the dust properties at each position on the disk.
That is, we can derive the spatially resolved ``snow line'' by this
method.\footnote{
The radiation from the hot dust ($\sim1000$ K) at the
disk inner edge is also scattered toward the observer as well as the
stellar radiation. This component may make a fraction of the NIR
scattered radiation from the disks. This effect could be taken into
account by modulating the incident spectrum which has been assumed to be
a Planck function (for example in eq.~[3]). Note that the modulation of
the incident spectrum results in only the change of the reference
color. In other words, we could use Figs.~4 and 5 even for such a case if
we take a preferable reference color rather than the Rayleigh-Jeans law.
}

Before moving the next section, we briefly comment on the effect of the
anisotropic scattering. If the grain size is enough large relative to
the wavelength, the scattering becomes a forward scattering not
isotropic. In terms of the asymmetry parameter 
$g \equiv \langle \cos \theta \rangle$ with the scattering angle
$\theta$, the NIR photons are scattered with $g>0.5$ by a 1 \micron\ 
silicate or H$_2$O ice grain. In this case, we expect that the disk
image in the scattered radiation shows asymmetry; if we observe an
inclined disk, the near side becomes brighter by the forward scattering
and the other side becomes fainter by an inefficient back scattering. 
On the other hand, the 3 \micron\ H$_2$O ice feature may not be affected
by the forward scattering because the asymmetry parameter does not show
a large variation around the feature.

\section{Observational feasibility of the ``snow line'' with current
 instruments}

\subsection{Required spatial resolution}

H$_2$O ice condenses on the surface of the dust grains/aggregates if the
temperature of the dust surface is low enough (100--200 K 
depending on the gas density; e.g., \cite{pol94}).
We estimate the ice existence region in optically thin disks and 
on the ``surface'' of optically thick disks.
For optically thick disks, only their ``surface'' is interesting 
because we cannot observe the scattered stellar radiation from the
optically thick interior.
When the disk height $z \ll R$, the temperature ($T_{\rm d}$) 
at the ``surface'' of optically thick disks or in optically thin disks 
is approximated to  
\begin{equation}
 \int_0^\infty \kappa_\nu^{\rm a} B_\nu(T_{\rm d}[R]) d\nu
  \approx \frac{1}{4}\left(\frac{R_*}{R}\right)^2 
  \int_0^\infty \kappa_\nu^{\rm a} B_\nu(T_*) d\nu\,,
\end{equation}
where we have neglected self-absorption of the disk radiation.

The dust temperature calculated from equation (12) is 
roughly proportional to $R^{-0.5}$ (e.g., \cite{hay85}). The absolute
value of the temperature depends on the dust composition and size 
as well as the central stellar luminosity. However, 
an order of estimate would be enough here.
When we assume the stellar effective temperature of 10,000 K 
and the stellar radius of 2.5 $\RO$ for an A-type (Herbig Ae/Be) star, 
and 3,000 K and 2.0 $\RO$ for a K-type (T Tauri) star, 
we find that the dust temperature becomes well below of 100 K 
if $R\gtrsim100$ AU for the A-type case and $R\gtrsim10$ AU for 
the K-type case, implying the presence of ``snow line'' around that
radius.

In order to resolve the disk radius of 10--100 AU at 100 pc away from us, 
we need to detect the disk scattered light as close as 0.1--1.0 arcsec
to the central star. 
At the same time, we must avoid the central stellar radiation 
to detect an image of the surrounding very faint scattered light from the
disk. Around 3 \micron\ (i.e. H$_2$O feature), a coronagraphic 
instrument with an adaptive optics on a 8-m telescope such as
CIAO (Coronagraphic Imager with Adaptive Optics) on the 8.2m Subaru
Telescope enables us to detect the disk scattered light as close as
about 0.75 arcsec \citep{fuk06}. It is still too large to resolve the
expected radius of the ``snow line'' for the K-type (T Tauri) case 
($R\gtrsim0.1$ arcsec), but is small enough to resolve the ``snow line''
for the A-type (Herbig) case ($R\gtrsim1$ arcsec).
Thus, we have a chance to detect the ``snow line'' on the disk around an
A-type star with the scattered light.

\subsection{Expected brightness}

An approximated form of the equation (9) is 
\begin{eqnarray}
 \frac{I^{\rm obs}_\nu}{\rm Jy~arcsec^{-2}}
 \approx0.076\left(\frac{T_*}{10^4~{\rm K}}\right)
 \left(\frac{\lambda}{3.8~\micron}\right)^{-2}
 \left(\frac{R_*}{2.5~\RO}\right)^2
 \left(\frac{R}{100~{\rm AU}}\right)^{-2} \cr
 \times \cases{ 
  \beta \omega_\nu & (optically thick disk) \cr
  \tau^{\rm s}_\nu & (optically thin disk) \cr} \,,
\end{eqnarray}
where we have assumed the Rayleigh-Jeans law, the visible fraction
$f_{\rm vis}=0.5$, and the $H$-function $H=1$. The spectra of the
brightness calculated by this equation are shown as the dashed curves in
Fig.~3; the good agreements ensure the validity of this approximated
equation.

We expect to detect the scattered light even in $L'$-band (3.8 \micron) 
from optically thick (i.e. protoplanetary) disks easily 
if the observing instrument can resolve the disks at $R\gtrsim100$ AU
with occulting the central star by a coronagraph. 
For example, we expect that the NIR brightness at 100 AU of an optically
thick disk around an A-type star is about 12 mag arcsec$^{-2}$ for a
scattering albedo $\omega\approx1$ (the dust size of $\gtrsim1$ \micron;
see Fig.~2) and a grazing angle $\beta=0.05$. This is consistent with
the observations of Herbig Ae disks (11--13 mag arcsec$^{-2}$; e.g., 
\cite{aug01,fuk04,fuk06}). Even for the case of a K-type star, 
the expected brightness at 100 AU is about 14 mag arcsec$^{-2}$ for
$\omega \approx 1$ and $\beta=0.05$. This is also consistent with the
observations of T Tauri disks (14--16 mag arcsec$^{-2}$; e.g., 
\cite{mcc02,wei02,sch03}).

Such brightness would be easily detected with a coronagraphic instrument
on a 8-m telescope. In order to clarify if ice exists on the surface of
the disk, we need to achieve $H_2O~(3~\micron) \gtrsim L' + 1$ mag 
(see Fig.~4 [c]). This would not be difficult because the sensitivity at
3 \micron\ is usually better than that at $L'$-band.

For an optically thin (i.e. debris) disk, we assume the scattering
optical depth of $\tau^{\rm s}_\nu = 10^{-3}$ 
which is based on the dust mass estimation in debris disks with the
SCUBA \citep{wil06}. 
In this case, the expected NIR brightness at 100 AU of the disk is about
16 mag arcsec$^{-2}$ for an A-type star. 
This is consistent with the observations of debris disks (15--16 mag 
arcsec$^{-2}$; e.g., \cite{gol06,sch06}). For a K-type star, the
expected brightness is about 18 mag arcsec$^{-2}$, which is also
consistent with that of AU Mic \citep{met05}. Thus, detecting debris
disks in $L'$ band may not be easy with ground-based telescopes even for
an A-type star case. With a space instrument, the spatial resolution is
the matter rather than the sensitivity.

\section{Summary}

We have proposed a new method for obtaining the spatially resolved 
``snow line'' in the circumstellar disks; 
observing 3 \micron\ H$_2$O ice feature in the scattered light.
Based on radiative transfer considerations, 
we show that the frequency dependence of the scattered light 
comes from the frequency dependence of the albedo for optically 
thick disks and of the scattering cross section for optically 
thin disks. Since 3 \micron\ H$_2$O ice feature is clearly imprinted 
in the albedo and the scattering cross section, we expect to find 
the feature in the scattered light easily. Then, we have proposed 
a diagnostics of disk dust properties by near-infrared 
three bands ($K$, $H_2O$, and $L'$) photometry; 
on the $K-H_2O$ and $K-L'$ two color diagram, 
we can distinguish that the disk is optically thick or thin, 
typical grain size is small or large, and ice exists or not.
Finally, we have confirmed the scattered light and the H$_2$O 
ice feature from protoplanetary disks 
are detectable and spatially resolvable 
with a current coronagraphic instrument on 8-m telescopes.

\bigskip

We are grateful to Dr. M.~Fukagawa for providing us with informations of 
the recent brightness measurements of protoplanetary and debris disks. 
We are also grateful to Drs. M.~Ishii, H.~Terada, N.~Takato,
Y.~K.~Okamoto and H.~Kawakita for useful comments.

\clearpage

\begin{figure}
 \begin{center}
  \FigureFile(120mm,100mm){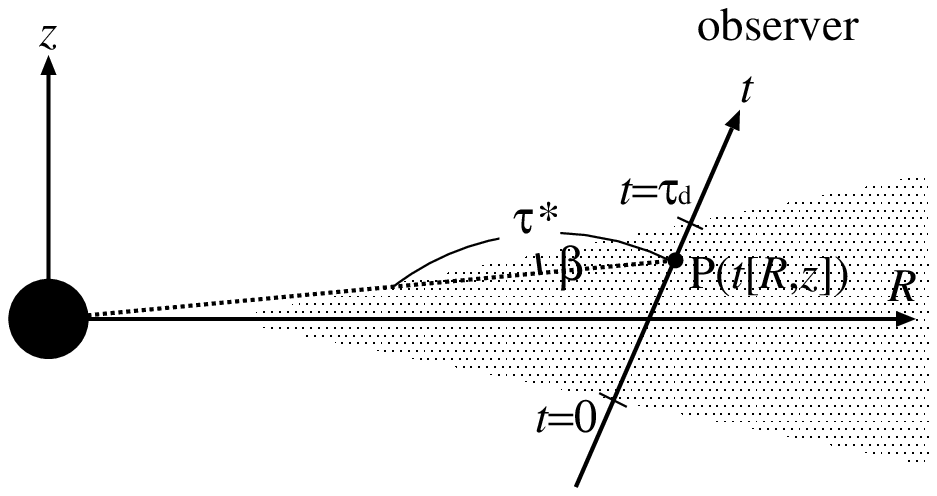}
 \end{center}
 \caption{Schematic figure of the considered geometry.}
\end{figure}

\clearpage

\begin{figure}
 \begin{center}
  \FigureFile(150mm,150mm){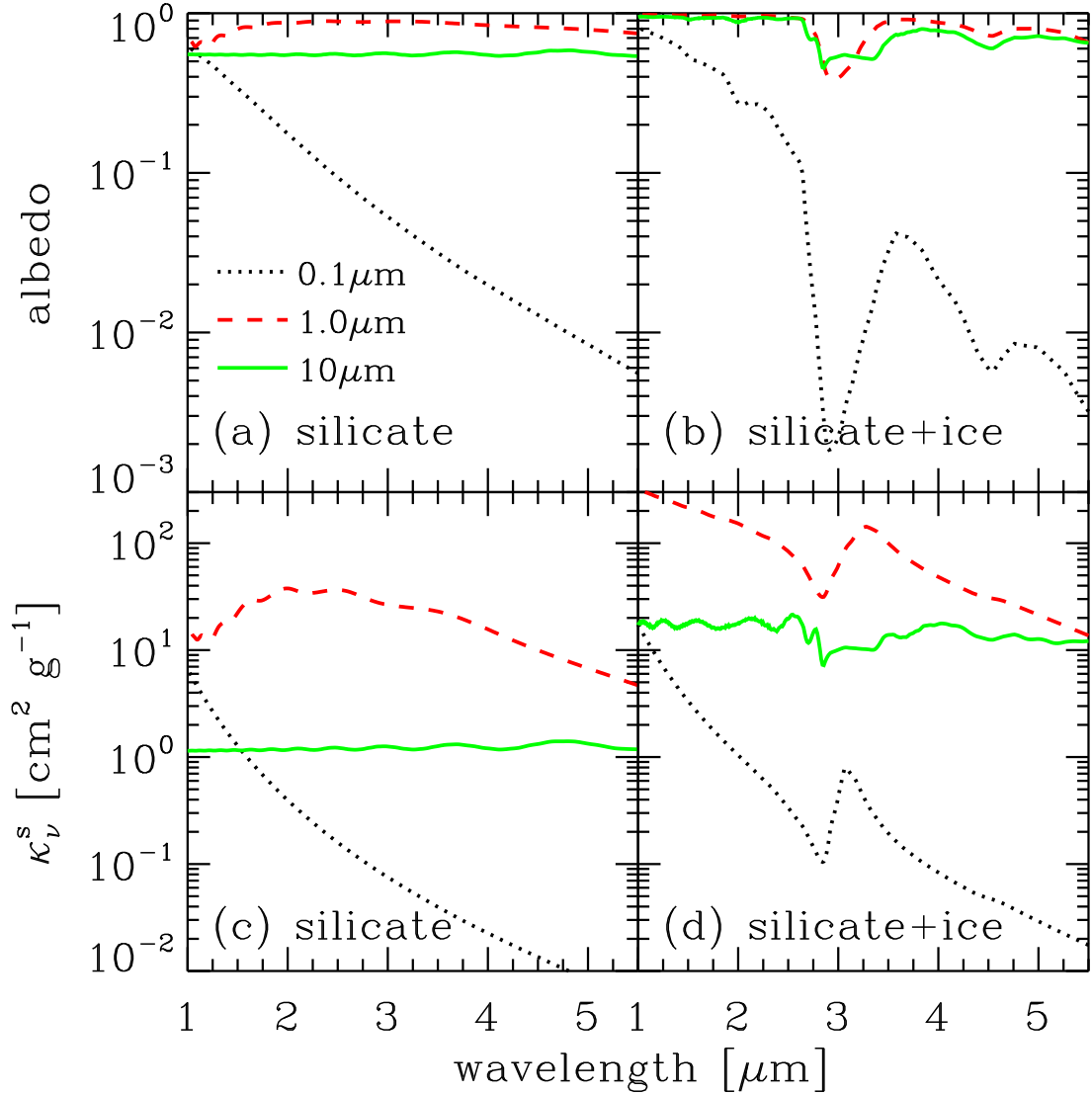}
 \end{center}
 \caption{Scattering albedos and cross sections: (a) albedo of 
 silicate dust, (b) albedo of silicate and H$_2$O ice dust, (c) cross  
 section per unit gas mass of silicate dust, and (d) cross 
 section per unit gas mass of silicate and H$_2$O ice dust. 
 The assumed mass abundances of silicate and H$_2$O ice relative
 to gas are 0.0043 and 0.0094, respectively (Miyake \& Nakagawa 1993). 
 The solid, dashed, and dotted lines are the cases with the grain size 
 of 10, 1, and 0.1 \micron, respectively.}
\end{figure}

\clearpage

\begin{figure}
 \begin{center}
  \FigureFile(150mm,150mm){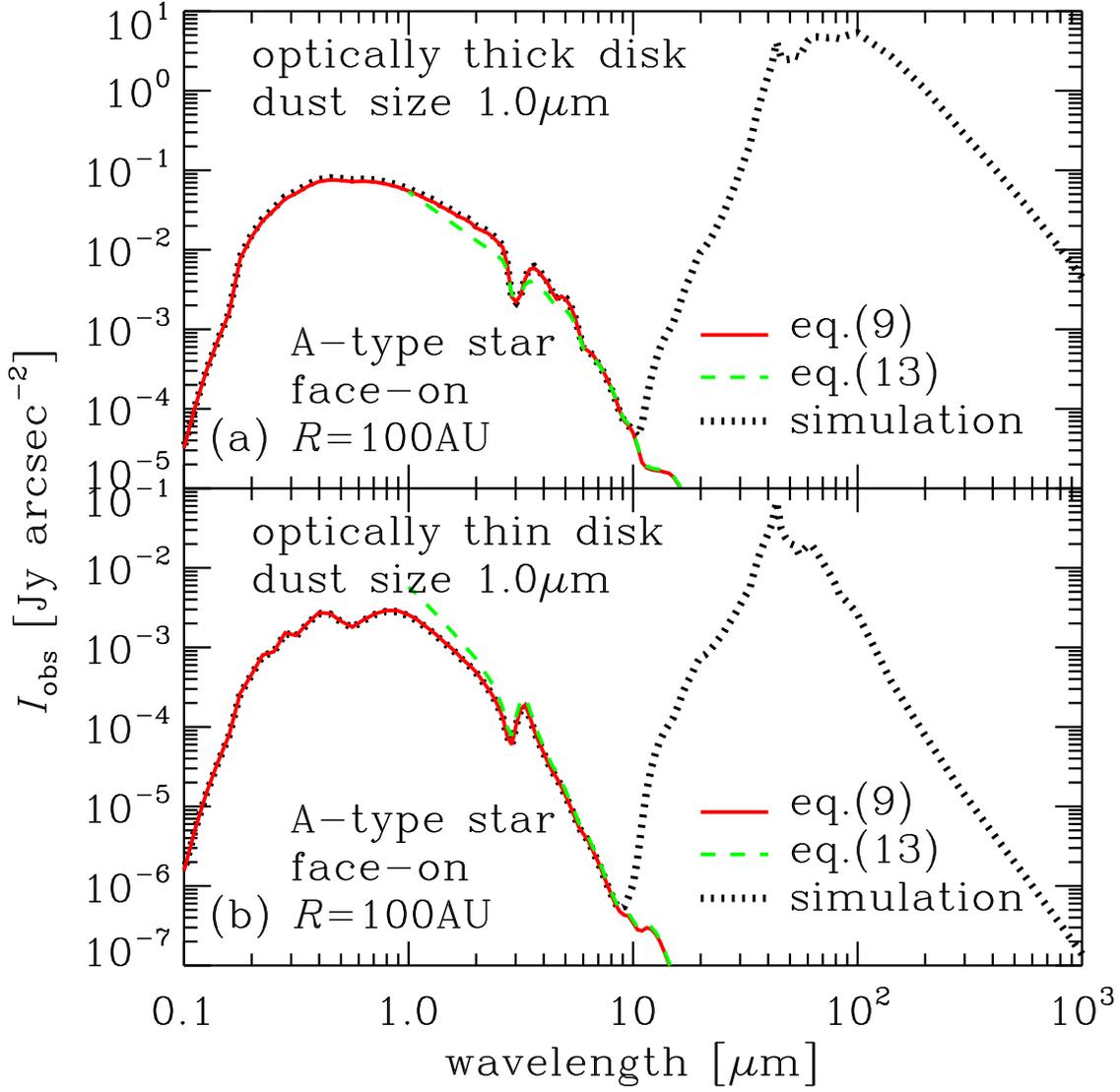}
 \end{center}
 \caption{Expected spectra of the face-on brightness of an annulus with
 the radius 100 AU around a central A-type star: (a) optically
 thick case and (b) optically thin case. The assumed dust is 1.0 \micron\ 
 silicate and H$_2$O ice. The solid curves are analytic models of the
 scattered light expressed in eq.~(9), the dashed curves are simpler 
 models of the scattered light expressed in eq.~(13), and the dotted
 curves are the scattered and the thermal radiations of the annulus from
 numerical radiative transfer simulations.}
\end{figure}

\clearpage

\begin{figure}
 \begin{center}
  \FigureFile(150mm,150mm){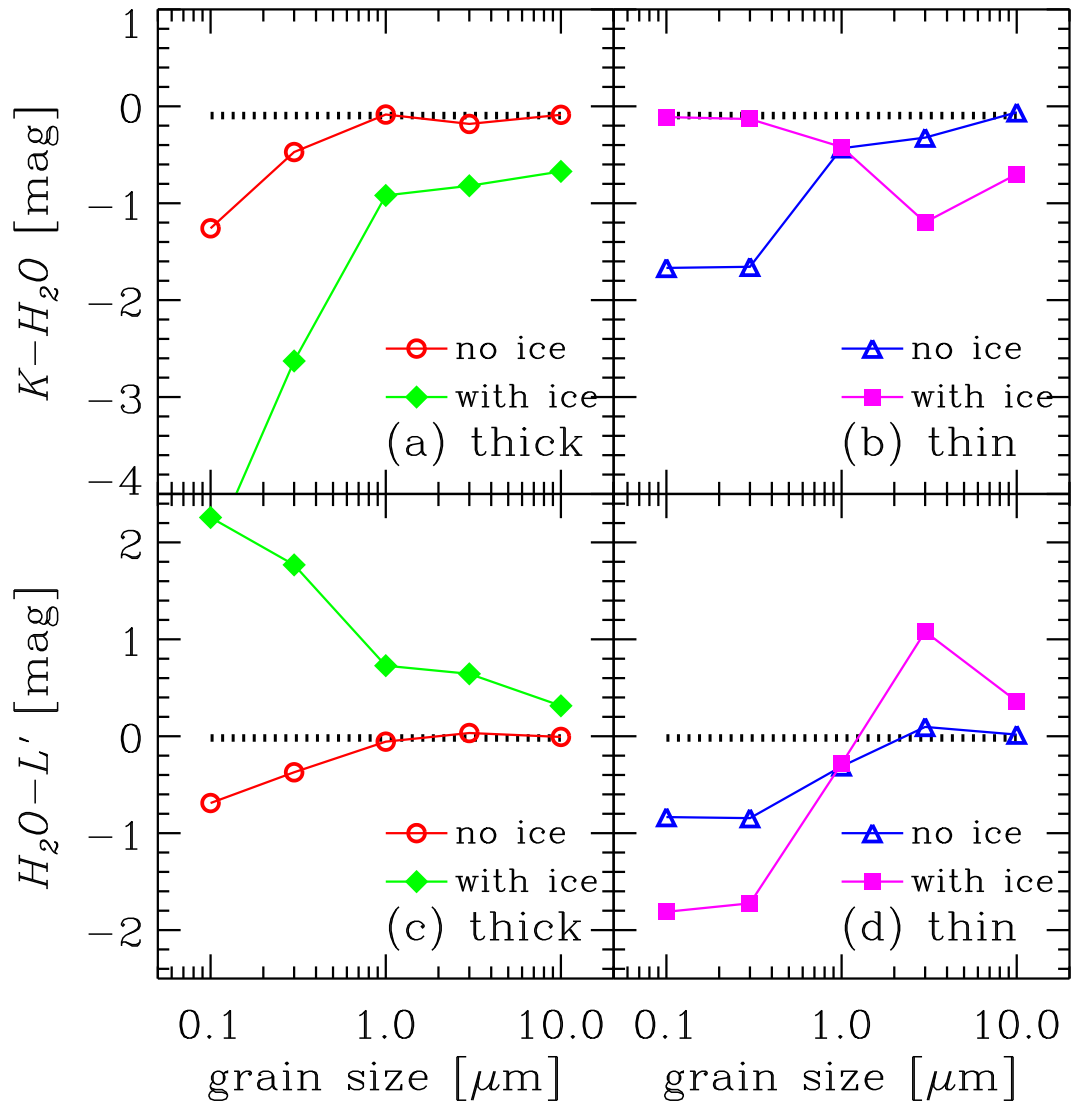}
 \end{center}
 \caption{Expected colors of the scattered light as a function of the
 dust size: (a) $K-H_2O$ for optically thick case, (b) $K-H_2O$ for
 optically thin case, (c) $H_2O-L'$ for optically thick case, and (d)
 $H_2O-L'$ for optically thin case. In each panel, open symbols are
 silicate dust and filled symbols are silicate and ice dust. The dotted
 line is the Rayleigh-Jeans law. The unit of the vertical axis is the
 Vega magnitude.}
\end{figure}

\clearpage

\begin{figure}
 \begin{center}
  \FigureFile(100mm,100mm){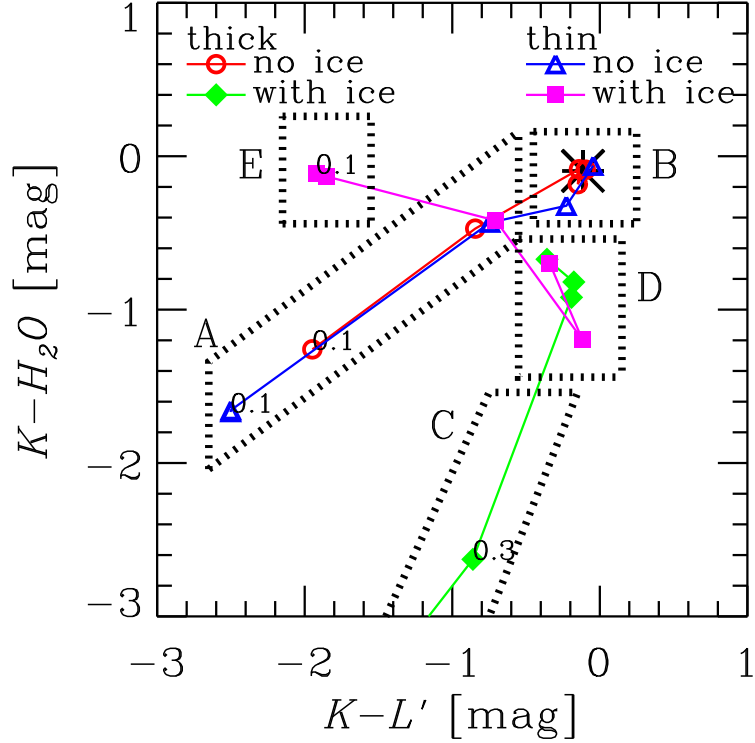}
 \end{center}
 \caption{Expected two color diagram. The circles, diamonds, triangles,
 and squares are the cases of the optically thick disk with silicate,
 the optically thick disk with silicate and ice, the optically thin disk
 with silicate, and the optically thin disk with silicate and ice,
 respectively. Each sequence with the same symbol is the sequence of the
 grain size: 0.1, 0.3, 1, 3, and 10 \micron. The labels on some points
 indicate the dust size in \micron. The triangles for 0.1 and 0.3
 \micron\ are almost overlaid. The large asterisk is the location 
 of the Rayleigh-Jeans law. The areas surrounded by dotted lines 
 show the five classifications on the diagram (see the text). The units
 of the axes are the Vega magnitude.}
\end{figure}

\end{document}